\documentclass[pre,showpacs]{revtex4}
\usepackage{makeidx}
\usepackage{graphicx}

\begin{document}

\title{Moving and colliding pulses in the subcritical Ginzburg-Landau model
with a standing-wave drive}
\author{Bakhtiyor B. Baizakov$^1$, Giovanni Filatrella$^2$, and Boris A.
Malomed$^3$}
\affiliation{$^1$ Physical-Technical Institute of the Uzbek Academy of Sciences, 2-b, G.
Mavlyanov str., 700084, Tashkent, Uzbekistan \\
$^2$ CNR-INFM, Laboratorio Regionale di Salerno and Dipartimento di Scienze
Biologiche ed Ambientali, Universit\`{a} del Sannio, via Port'Arsa 11, 82100
Benevento, Italy \\
$^{3}$ Department of Interdisciplinary Studies, School of Electrical
Engineering, Faculty of Engineering, Tel Aviv University, Tel Aviv 69978,
Israel }

\begin{abstract}
We show the existence of steadily moving solitary pulses (SPs) in
the complex Ginzburg-Landau (CGL) equation, which includes the
cubic-quintic (CQ) nonlinearity and a conservative linear driving
term, whose amplitude is a standing wave with wavenumber $k$ and
frequency $\omega $, the motion of the SPs being possible at
velocities $\pm \omega /k$, which provide locking to the drive. A
realization of the model may be provided by traveling-wave
convection in a narrow channel with a standing wave excited in its
bottom (or on the surface). An analytical approximation is
developed, based on an effective equation of motion for the SP
coordinate. Direct simulations demonstrate that the effective
equation accurately predicts characteristics of the driven motion
of pulses, such as a threshold value of the drive's amplitude.
Collisions between two solitons traveling in opposite directions
are studied by means of direct simulations, which reveal that they
restore their original shapes and velocity after the collision
\end{abstract}

\maketitle

\section{Introduction and the model}

Complex Ginzburg-Landau (CGL) equations are universal models describing
formation of extended patterns and solitary pulses (SPs) in dissipative
nonlinear media driven by intrinsic gain. These models are interesting by
themselves \cite{reviews2}, and also due to their various physical
applications. In particular, SP solutions explain the existence of localized
pulses observed in traveling-wave convection (TWC) \cite{Kolodner}, as well
as pulsed (soliton) operation regimes featured by fiber lasers \cite{laser}.

Exact SP solutions are available in the CGL equation with the cubic
nonlinearity \cite{Lennart}, but these pulses are unstable. A
straightforward generalization of the model, which makes it possible to
create stable SPs, is provided by the cubic-quintic (CQ) nonlinearity.
The accordingly modified CGL equation includes linear loss and cubic
gain (on the contrary to the linear gain and cubic loss in the cubic
equation), and an additional quintic lossy term that provides for the
overall stability. The CQ CGL equation was introduced by Petviashvili
and Sergeev (in fact, in two dimensions) \cite{NizhnyNovgorod}, and its
stable SP solutions (in one dimension) were first predicted, using an
analytical approximation based on the proximity to the nonlinear
Schr\"{o}dinger (NLS)\ equation, in Ref. \cite{PhysicaD}. Later, pulses
and their stability in the CQ CGL equation were investigated in detail
\cite{CQCGL,Hidetsugu}.

It is relevant to note that stable SPs may also be found in a model
with the cubic nonlinearity only, which is based on the cubic CGL
equation linearly coupled to an extra linear dissipative equation
\cite{Herb-Javid}. This system gives rise to exact analytical solutions
for \emph{stable} SPs \cite{Javid}, and to (numerically found)
breathers, i.e., randomly oscillating and \emph{randomly walking}
robust pulses \cite{Hidetsugu}. In this connection, it is necessary to
mention that standing chaotic pulses were found in the CQ\ CGL equation
too \cite{DeisslerBrandChaotic}, and such pulses were observed
experimentally in electrohydrodynamic convection in liquid crystals
\cite{ExperimentElectroConvChaoticPulse}.

A problem of fundamental interest is the understanding of conditions
providing for the existence of steadily moving SPs. At the model level,
the motion of pulses at arbitrary velocity (and collisions between
them) are obviously possible in the CQ CGL equation that does not
include a diffusion term \cite{HS}. In the experiment, pulses
persistently moving in either direction, at a very low uniquely
determined velocity, were observed in TWC \cite{Kolodner}. The motion
and its extreme slowness were explained by coupling between the CGL
equations for the right- left-traveling waves (which can be replaced by
a single fourth-order CGL equation \cite{Zeitschrift}) and an
additional real diffusion-type equation for the concentration field in
a binary fluid, where the TWC experiments are performed \cite{Riecke}.

SPs may be set in persistent motion not only by their intrinsic dynamics,
but also by an external drive. In physically relevant situations, this
includes a combination of a spatially periodic inhomogeneity -- typically,
in the form of $\cos (kx)$ -- and a time-periodic (ac) driving field applied
to the system, in the form of $\cos (\omega t)$. Although the two factors
usually act separately (i.e., they appear in different terms of the
corresponding equation), the interplay between them naturally suggest a
possibility to observe motion of pulses at fundamental \textit{resonant
velocities}, $\pm \omega /k$, and, possibly, also at velocities
corresponding to higher spatial and temporal harmonics, i.e., $\pm (m\omega
/nk)$, with integer $m$ and $n$.

In models (different from the CGL equations) which give rise to
\textit{topological solitons}, the ac-driven progressive motion of
solitons in dissipative media was predicted in both discrete settings
(in which case the discreteness itself provides for the periodic
spatial inhomogeneity), based on the Toda \cite{TL} and
Frenkel-Kontorova \cite{FK} lattices, and in continuum models based on
modified sine-Gordon equations, that describe weakly damped
periodically inhomogeneous ac-driven long Josephson junctions (LJJs)
\cite{LJJ}. Supporting motion of a topological soliton by the ac drive
is relatively easy, as the driving field directly couples to the
topological charge [in particular, the bias current applied to LJJ
directly acts on the magnetic-flux quantum carried by the corresponding
topological soliton (fluxon)]. For this reason, the motion of the
fluxon in a \emph{circular} LJJ may also be supported by a
\textit{rotating magnetic field} (i.e., as a matter of fact, by an
external traveling wave) applied in the plane of the junction
\cite{rotating}. In all these cases, the ac-driven motion is possible
if the amplitude of driving field, $\epsilon $, exceeds a certain
threshold value, $\epsilon _{\mathrm{thr}}$ (with $\epsilon <\epsilon
_{\mathrm{thr}}$, the traveling-wave drive cannot support the motion of
a topological soliton at the velocity equal to the phase velocity of
the wave, but it can \textit{drag} the soliton, supporting its motion
in a \textit{slow-drift} regime \cite{drift}).

The predicted modes of the ac-propelled motion of solitons at the
resonant velocities were observed experimentally, both in a
discrete electric (LC) transmission lattice \cite{TL}, and in a
circular LJJ \cite{Ustinov}. It is relevant to stress that all the
above-mentioned regimes of the ac-driven motion are possible in
either direction, hence they are completely different from various
\textit{soliton ratchets}, that have been studied in detail
theoretically and experimentally \cite{ratchets}.

A challenging problem is to find mechanisms making it possible to support
persistent motion of \emph{nontopological} (dynamical) SPs in dissipative
media (such as the pulses in CGL models), where the driving force cannot
move them directly. Thus far, this was only reported for discrete solitons
in a weakly damped Ablowitz-Ladik lattice driven by an ac field \cite{AL}.

In this work, we aim to demonstrate that such a stable dynamical regime can
be found in what may be considered as a paradigmatic model, viz., the CQ CGL
equation with an extra conservative term representing a standing wave acting
on the system:
\begin{equation}
iu_{t}+\frac{1}{2}u_{xx}+|u|^{2}u=-i\alpha u+i\beta u_{xx}+i\gamma
|u|^{2}u-i\Gamma |u|^{4}u-\epsilon \cos \left( kx\right) \cos \left( \omega
t\right) u,  \label{GL}
\end{equation}
Coefficients $\alpha \geq 0,\beta \geq 0$ and $\gamma >0,\Gamma >0$ on the
right-hand side of Eq. (\ref{GL}) account for the linear loss, effective
diffusion, cubic gain, and quintic loss, respectively. Using the scale
invariance of Eq. (\ref{GL}), we will fix $k\equiv 1/4$, while $\epsilon $
and $\omega $ remain arbitrary parameters.

A straightforward physical realization of Eq. (\ref{GL}) can be found
in terms of TWC (in a binary fluid), which, as said above, is
adequately modeled (in the region of the subcritical instability) by
the CQ CGL \cite{Kolodner}. The driving term, which is a conservative
one [i.e., it yields zero contribution to the balance equation for the
norm, see Eq. (\ref{M}) below], may be generated by a standing elastic
(acoustic) wave created on the bottom of the convection cell (or,
possibly, a standing wave created on the surface of the convection
layer), provided that the drive's wavelength, $2\pi /k$, is much larger
than the critical wavelength of the convection-triggering instability,
and $\omega $ must be sufficiently small too. The latter conditions
comply with the analysis developed below, which is focused on the case
of $k/\eta \ll 1$, and needs to have $\omega $ small enough too, to
provide for a reasonably small locking threshold, see Eqs. (\ref{thr})
and (\ref{small-k-thr}) below.

The coefficient in front of the driving term in Eq. (\ref{GL}) can be
decomposed into a superposition of two counter-propagating traveling waves,
\begin{equation}
\cos \left( kx\right) \cos \left( \omega t\right) =\frac{1}{2}\left[ \cos
\left( k\left( x-c_{0}t\right) \right) +\cos \left( k\left( x+c_{0}t\right)
\right) \right] ,  \label{traveling}
\end{equation}

\noindent where $c_{0}\equiv \omega /k$. As said above, this decomposition
suggests that the drive may support motion of a soliton with either positive
or negative resonant velocity, $c=\pm c_{0}$. However, the actual existence
of stable moving solitons, locked to either traveling-wave component of the
drive, is not obvious. An objective of this work is to predict this
possibility in an analytical approximation, and verify it by numerical
simulations of Eq.(\ref{GL}), identifying a parameter region in which SPs
can steadily move at the resonant velocity (slow drift of driven pulses is
formally possible too, but it turns out to be always unstable, in the
present model).

The analytical and numerical findings are presented below in Sections II and
III, respectively. In concluding Section IV, we summarize the obtained
results.

\section{Analytical approximation}

Equation (\ref{GL}) is written in the form of a perturbed
nonlinear Schr\"{o}dinger (NLS) equation, which, in the absence of
perturbations ($\alpha =\beta =\gamma =\Gamma =\epsilon =0$), has
a family of ordinary soliton solutions,
\begin{equation}
u_{\mathrm{sol}}=\eta ~\mathrm{sech}\left( \eta (\left[ x-\xi
(t)\right] \right) ~e^{icx+(i/2)\left( \eta ^{2}-c^{2}\right) t},\quad
~\dot{\xi}=c, \label{soliton}
\end{equation}
with arbitrary amplitude $\eta $ and velocity $c$ (the overdot stands
for $d/dt$). For $\epsilon =0$ (no driving term), but with $\alpha
,\beta ,\gamma ,\Gamma >0$, two stationary solitons are selected from
continuous family (\ref{soliton}) by the norm-balance condition,
\begin{equation}
\frac{d}{dt}\left( \int_{-\infty }^{+\infty }|u(x)|^{2}dx\right) =0.
\label{M}
\end{equation}
Within the framework of the perturbation theory, Eq. (\ref{M}) yields
the following values of the amplitude \cite{PhysicaD}:
\begin{equation}
\eta _{\pm }^{2}=\frac{5\left( 2\gamma -\beta \right) \pm \sqrt{5\left[
5\left( 2\gamma -\beta \right) ^{2}-96\alpha \Gamma \right] }}{16\Gamma },
\label{eta}
\end{equation}
while the velocity is zero. The pulses corresponding to larger and smaller
values given by Eq. (\ref{eta}), i.e., $\eta _{+}$ and $\eta _{-}$, are
stable and unstable, respectively.

To proceed with the analytical approach for $\epsilon \neq 0$, we make
use of the balance equation for the momentum, $P=i\int_{-\infty
}^{+\infty }u_{x}^{\ast }udx$, with asterisk standing for the complex
conjugate. Indeed, an exact corollary of Eq. (\ref{GL}) is
\[
\frac{dP}{dt}=-2\alpha P-\beta \int_{-\infty }^{+\infty }i\left(
u_{x}u_{xx}^{\ast }-u_{x}^{\ast }u_{xx}\right) dx
\]
\begin{equation}
+\int_{-\infty }^{+\infty }\left( \gamma |u|^{2}-\Gamma |u|^{4}\right)
i\left( u_{x}u_{xx}^{\ast }-u_{x}^{\ast }u_{xx}\right) dx-\epsilon k\cos
(\omega t)\int_{-\infty }^{+\infty }\sin (2kx)\cdot |u|^{2}dx~.  \label{P}
\end{equation}
In the spirit of the multiscale perturbation theory, we assume that the loss
and gain terms in Eq. (\ref{GL}) represent stronger perturbations, hence
they select value (\ref{eta}), with the upper sign, for $\eta $. Strictly
speaking, the use of that expression implies that the velocity is small
enough, $\dot{\xi}^{2}\ll \eta _{+}^{2}$ [then, an additional contribution
to the norm-balance equation (\ref{M}) for the moving solitons, $\sim \beta
\dot{\xi}^{2}$, may be neglected]. After that, the substitution of
expression (\ref{soliton}) with the constant amplitude, $\eta =\eta _{+}$,
in Eq. (\ref{P}) and straightforward integrations yield the following
effective equation of motion for the central coordinate, $\xi (t)$ (we omit
the subscript in $\eta _{+}$):
\begin{equation}
\ddot{\xi}=-\frac{4}{3}\beta \eta ^{2}\dot{\xi}-2\beta
\dot{\xi}^{3}-\frac{\pi \epsilon k^{2}}{2\eta \sinh \left( \pi k/2\eta
\right) }\sin \left( k\xi \right) \cos (\omega t).  \label{xi}
\end{equation}
\noindent In the case of narrow solitons, $k/\eta \ll 1$, Eq. (\ref{xi})
simplifies:
\begin{equation}
\ddot{\xi}=-\frac{4}{3}\beta \eta ^{2}\dot{\xi}-2\beta \dot{\xi}
^{3}-\epsilon k\sin \left( k\xi \right) \cos (\omega t).
\label{small-k}
\end{equation}
\noindent Equation (\ref{small-k}) with $\epsilon =0$ (no drive) admits an
explicit analytical solution:
\[
\dot{\xi}^{2}(t)=\frac{2}{3}\frac{\eta ^{2}}{\left( 1+2\eta
^{2}/3\dot{\xi}_{0}^{2}\right) \exp \left( 3t/2\beta \eta ^{2}\right)
-1}.
\]

In the presence of the drive, motion at the average velocity $c_{0}$ implies
the existence of a solution to Eq. (\ref{xi}) in the form of $\xi
(t)=c_{0}\left( t-t_{0}\right) +\Xi (t)$, with some constant $t_{0}$ and
time-periodic terms $\Xi (t)$. Applying to Eq. (\ref{xi}) known methods used
in the analysis of the ac-driven motion of solitons at the resonant velocity
\cite{TL} - \cite{rotating}, we conclude that, in the present case, the
ac-propelled resonant motion may be possible, provided that the drive's
amplitude exceeds a threshold value,
\begin{equation}
\epsilon >\epsilon _{\mathrm{thr}}=\frac{8\beta \omega \eta \left(
2\eta ^{2}k^{2}+3\omega ^{2}\right) }{3\pi k^{5}}\sinh \left( \frac{\pi
k}{2\eta } \right).   \label{thr}
\end{equation}

\noindent Actually, the threshold is identified as the smallest value
of $\epsilon $ at which the average value of the last (driving) term on
the right-hand side of Eq. (\ref{xi}) may compensate the first two
(braking) terms, with $\dot{\xi}=\omega /k$. For narrow solitons, with
$k/\eta \ll 1$, expression (\ref{thr}) simplifies,
\begin{equation}
\epsilon >\epsilon _{\mathrm{thr}}=\frac{4\beta \omega }{3k^{4}}\left( 2\eta
^{2}k^{2}+3\omega ^{2}\right) .  \label{small-k-thr}
\end{equation}

As mentioned above, in the case of $\epsilon <\epsilon _{\mathrm{thr}}$ one
may consider a \textit{drift }regime, in which the SP is dragged by either
traveling wave from Eq. (\ref{traveling}) at a small velocity, $c\ll c_{0}$.
However, additional analytical considerations, as well as direct
simulations, demonstrate that the drift mode of motion is \emph{always
unstable} in the present system (unlike the resonant-motion regime, which
may definitely be stable).

\begin{figure}[tbp]
\centerline{\includegraphics[width=8cm,height=6cm,clip]{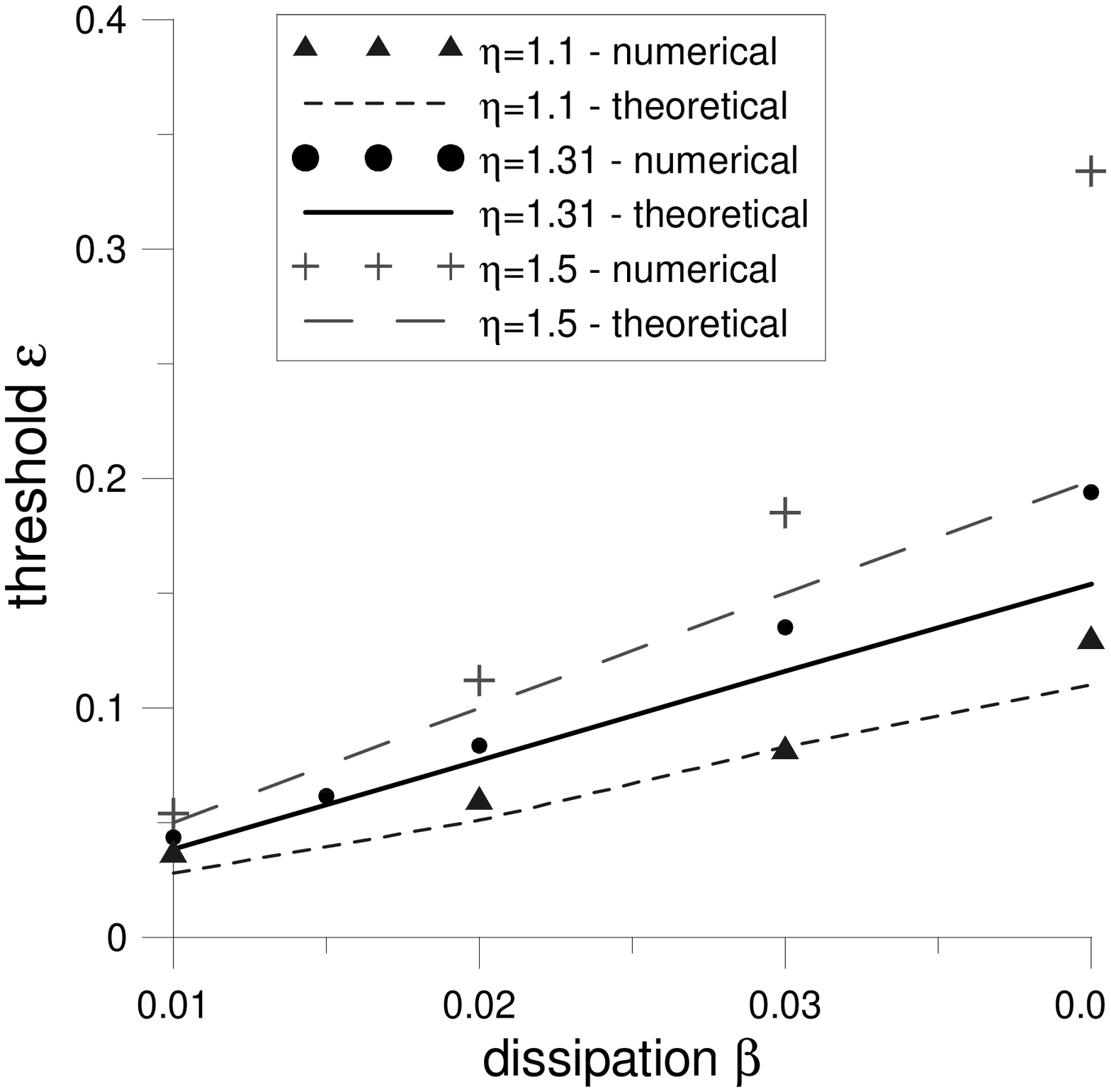}}
\caption{Comparison of the numerical results (symbols) and the
theoretical prediction, given by Eq. (\protect\ref{thr}) (lines), for
the minimum amplitude of the external drive $\protect\epsilon $
necessary to support a steady traveling solution in Eq.
(\protect\ref{xi}). Here and in all examples displayed below, the
driving wave is taken with $k=0.25$ and $\protect\omega =0.05$.}
\label{fig1}
\end{figure}

As shown in Fig. \ref{fig1}, we have compared analytical prediction
(\ref{thr}) for the threshold amplitude with findings following from
simulations of Eq. (\ref{xi}). The results are in good agreement for
low dissipation rates, and the agreement deteriorates with the increase
of $\beta \eta ^{2}$, which is quite natural, as the perturbation
theory is based on the assumption of weak dissipation. Generally,
analytical expression (\ref{thr}) underestimates the actual threshold.

\begin{figure}[tbp]
\centerline{\includegraphics[width=8cm,height=6cm,clip]{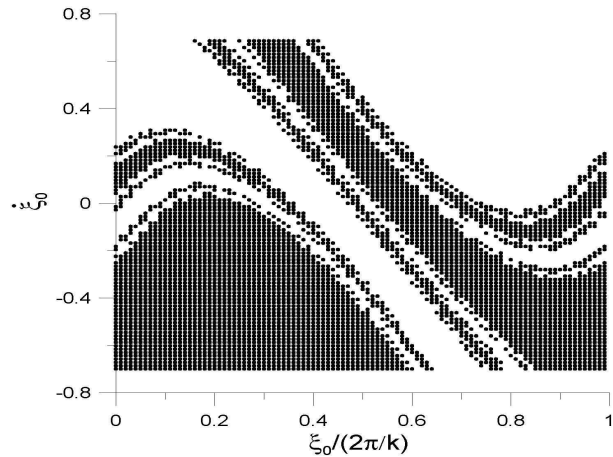}}
\caption{Results of the numerical integration of the equation of
motion, Eq. (\protect\ref{xi}), with $\beta=0.02$, $\epsilon=0.1$, and
$\eta=1.31$, obtained by varying initial values of the coordinate and
velocity, $\protect\xi _{0}$ and $\dot{\protect\xi}_{0}$. White space
and black dots denote, respectively, established states in the form of
the motion with resonance velocities $c_{0}=+0.2$ and $-0.2$,
respectively.} \label{fig2}
\end{figure}

A typical example of the dependence of the established ac-driven
regimes of motion [as per effective equation of motion (\ref{xi})] on
initial values of the coordinate and velocity is displayed in Fig.
\ref{fig2}. For the explored parameter range, simulations of Eq.
(\ref{xi}) demonstrate that the motion regimes corresponding to the
fundamental-resonance velocities, $\pm \omega /k$, are the single
stable states (\textit{attractors}) for $\epsilon
>\epsilon _{\mathrm{thr}}$ (velocities corresponding to the locking of a
moving SP to higher-order resonances do not emerge in the simulations),
while only the states with zero average velocity are stable for
$\epsilon <\epsilon _{\mathrm{thr}}$. Despite the interweaving
attraction basins of the two stable regimes, which is quite a notable
feature of Fig. \ref{fig2}, the simulations of Eq. (\ref{xi}) do not
reveal any hysteresis in the established states (i.e., no overlap
between the two attraction basins).

\section{Numerical simulations of the complex Ginzburg-Landau equation}

\subsection{Verification of the ac-driven regime of motion}

To verify and expand the predictions of the analytical approach
developed in the previous section, we have performed direct numerical
simulations of the underlying PDE, i.e., the driven CQ CGL equation,
Eq. (\ref{GL}). First, we have evaluated the amplitude of the soliton
for a given set of gain and loss parameters in the CQ CGL equation
(\ref{GL}) with $\epsilon =0$, using the above analytical predictions.
For instance, Eq. (\ref{eta}) with $\alpha =0.01$, $\beta =0.02$,
$\gamma =0.06$, and $\Gamma =0.03$ predicts the stable soliton's
amplitude, $\eta _{+}=1.31$. Next, we have estimated the threshold
strength, $\epsilon _{\mathrm{thr}}$, of the ac drive in Eq.
(\ref{GL}). Taking (as in the above examples) $k=0.25$ and $\omega
=0.05$, which correspond to resonance velocity $c_{0}\equiv \omega
/k=0.2$, Eq. (\ref{thr}) yields $\epsilon _{\mathrm{thr}}=0.077$. The
spatial period (i.e., $k$) was selected so as to minimize excitation of
internal vibrations in the SP under the action of the driving term. To
this end, narrow pulses were chosen, with $\eta k\ll 1$ [note this is
the same assumption which made it possible to simplify Eq. (\ref{xi}),
replacing it by Eq. (\ref{small-k}), and Eq. (\ref{thr}) by Eq.
(\ref{small-k-thr})].

\begin{figure}[tbp]
\centerline{\includegraphics[width=6cm,height=6cm,clip]{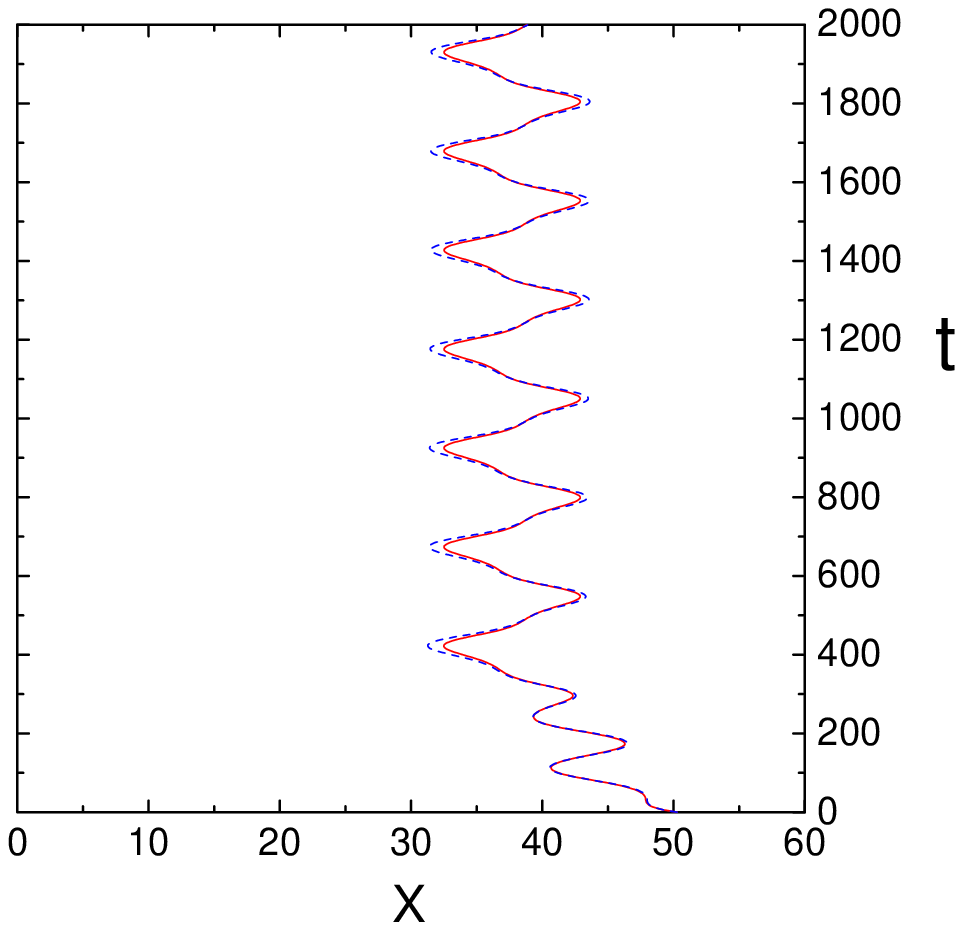}
      \quad \includegraphics[width=6cm,height=6cm,clip]{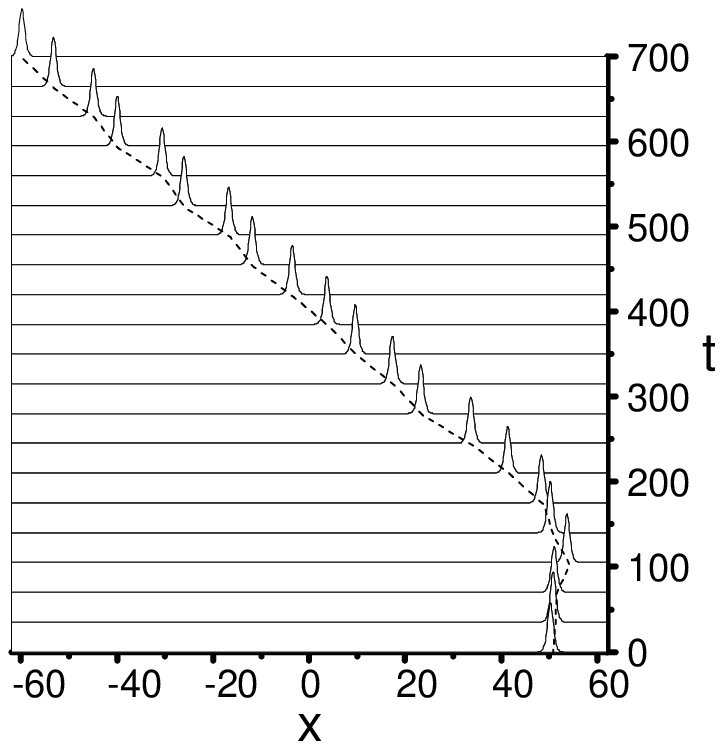}}
\caption{(Color online) Soliton with amplitude $\protect\eta =1.31$ is
set, at $t=0$, at point $\protect\xi _{0}=16\protect\pi $, and shoved
by lending it initial velocity $c_{0}=-0.2$ [in fact, it means
amplifying the soliton by $\exp (ic_{0}x)$]. The coefficients in Eq.
(\protect\ref{GL}) are $ \protect\alpha =0.01$, $\protect\beta =0.02$,
$\protect\gamma =0.06,$ $ \Gamma =0.03$, and $\protect\omega
=0.05,k=0.25$. Left panel: The motion of the soliton's center as
predicted by Eq. (\protect\ref{xi}) (blue dashed line), and as found
from the direct simulations of the full CGL equation, Eq.
(\protect\ref{GL}) (red continuous line). When the strength of the
driving force is smaller than the threshold value, e.g.,
$\protect\varepsilon =0.05\ $[recall Eq. (\protect\ref{thr}) predicts
$\protect\varepsilon_{\mathrm{thr}}=0.077$ for the present case],
consistent motion of the solitary pulse does not occur. Right panel:
When the strength of the driving force exceeds the threshold, e.g.,
$\protect\varepsilon =0.1>\protect\varepsilon _{\mathrm{thr}}$, the
soliton moves steadily, as expected. In the picture, we juxtaposed a
series of the soliton's profiles extracted from the numerical solution
of Eq. (\protect\ref{GL}), with the trajectory of motion predicted by
Eq. (\protect\ref{xi}) (dashed line). } \label{fig3}
\end{figure}

In Fig. \ref{fig3} we display the simulated evolution of the SP which
was initially placed at position $\xi _{0}=16\pi $ and set in motion
with the initial velocity $c_{0}=-0.2$. If the driving potential
strength is below the threshold value, $\epsilon <\epsilon
_{\mathrm{thr}}$, the SP gets trapped within a finite domain, as seen
in the left panel of Fig. \ref{fig3}; on the other hand, simulations of
Eq. (\ref{GL}) with $\epsilon >\epsilon _{\mathrm{thr}}$ reveal, as
expected, stable motion of the solitary pulse, see the right panel of
Fig. \ref{fig3}.

Systematic simulations of Eq. (\ref{GL}) corroborate the predictions of the
analytical approach based on Eq. (\ref{xi}). We did not aim to find the
threshold value of the drive's amplitude from the PDE simulations with a
very high accuracy, but, generally, the numerical results are in accordance
with the above prediction given by Eq. (\ref{thr}).

\subsection{Collisions between solitary pulses}

\begin{figure}[tbp]
\centerline{\includegraphics[width=8cm,height=6cm,clip]{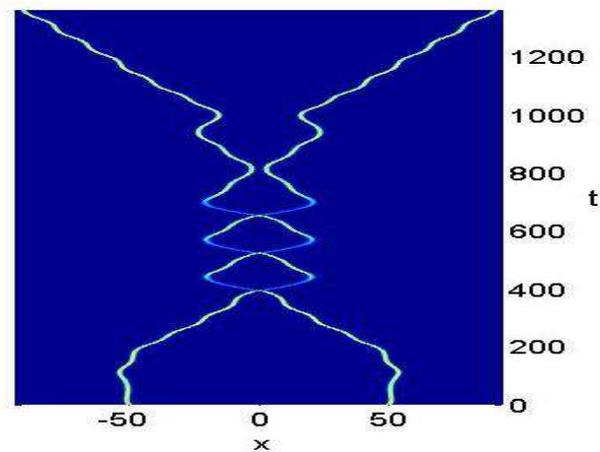}}
\caption{(Color online) Collision of two ac-driven solitary pulses with
opposite velocities, $c_{0}=\pm 0.2$. After a series of repeated
collisions, the pulses eventually separate, completely restoring their
initial shape and velocities. Parameters are the same as in the right
panel of Fig. \protect \ref{fig3}. } \label{fig4}
\end{figure}

The existence of stably moving ac-propelled SPs suggest a possibility
to consider collisions between them. Figure \ref{fig4} displays a
simulated collision of two identical pulses moving with opposite
velocities, $c_{0}=\pm 0.2$, both being driven by the standing wave
with $\epsilon =0.1$. This and other runs of the simulations
demonstrate that the solitons survive the collision. As seen from Fig.
\ref{fig4}, each soliton readily recovers its original shape and
velocity after the collision, which is easily explained by the fact
that these shapes and velocities correspond to unique attractors of the
ac-driven CQ\ CGL\ equation, see above. A generic feature obvious in
Fig. \ref{fig4} and observed in many other runs of simulations is that
\emph{multiple} collisions actually occur between the moving pulses.

\begin{figure}[tbp]
\centerline{\includegraphics[width=8cm,height=6cm,clip]{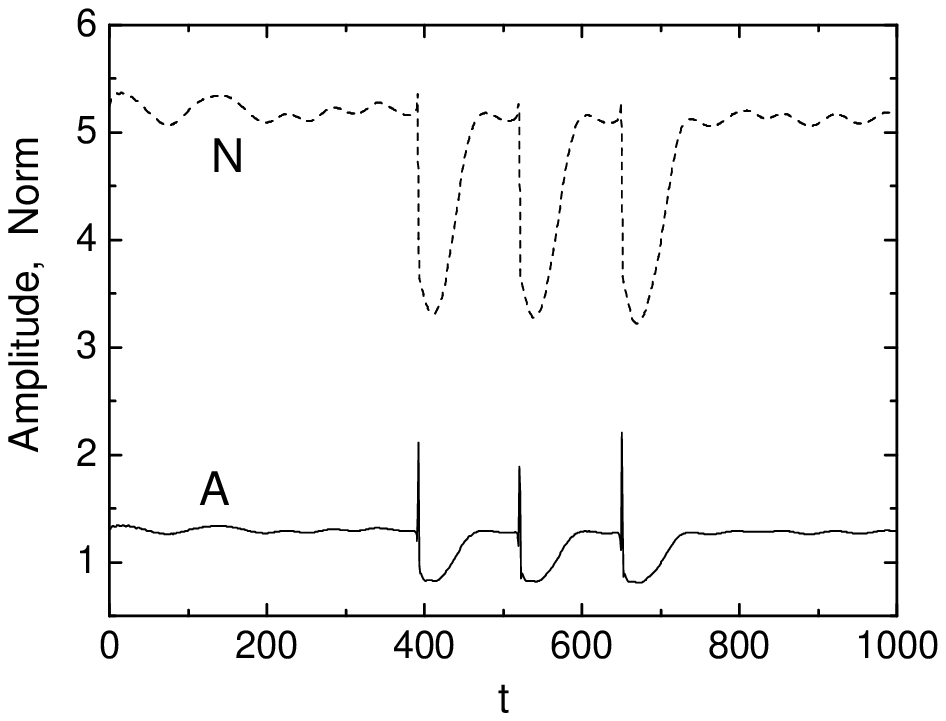}}
\caption{Time dependencies of the amplitude of the wave field ($A$) and
its total norm ($N$) additionally illustrating the collision between
two solitary pulses shown in Fig. \protect\ref{fig4}.} \label{fig5}
\end{figure}

The collisions are additionally illustrated, in Fig. \ref{fig5}, by
plots showing the norm and amplitude of the colliding pulses as
functions of time. It is seen that the total norm undergoes violent
changes when the collision takes place, but later it quickly recovers
the original value. No tangible radiation coming out from the collision
region have be found in the simulations.

\section{Conclusion}

In this paper, we have shown that solitary pulses in the cubic-quintic CGL
equation which contains the conservative standing-wave driving term can lock
to the resonant velocities provided by the drive, and thus move to right or
left in a stable fashion. This dynamical regime was predicted in an
analytical form, by means of the effective equation of motion for coordinate
of the solitary pulse, and corroborated by direct simulations of the full
CGL equation, with the driving term. In particular, the threshold value of
the drive's amplitude (the minimum amplitude necessary to lock the pulse to
the resonant velocity), which was predicted by the effective equation of
motion, is quite close to values provided by the full simulations. Unlike
the motion mode with the velocity corresponding to the fundamental resonance
with the external drive, motion of a pulse locked to a higher-order
resonance is never observed, nor is a dragging (slow-drift) mode ever found.
Collision between ac-driven pulses moving in opposite directions were also
studied, by means of the direct simulations. It was concluded that the
solitary pulses are robust in this sense too, recovering the initial shapes
and velocities after the collision.

\section*{Acknowledgements}

We thank Mario Salerno for valuable discussions. B.B.B. acknowledges
partial support from the Fund for Fundamental Research of the Uzbek
Academy of Sciences, grant No. 20-06. The work of B.A.M. was supported,
in a part, by the Israel Science Foundation through the
Center-of-Excellence grant No. 8006/03.

\end{document}